\documentclass[conference]{IEEEtran}
\IEEEoverridecommandlockouts

\usepackage{amssymb}
\usepackage{amsmath}
\usepackage{pifont}

\usepackage{hyperref}

\usepackage{todonotes}

\usepackage{graphicx}

\usepackage{verbatim}

\usepackage{mathrsfs}

\usepackage{enumitem}

\usepackage{array}

\usepackage{xspace}
\newcommand{\etal}{\textit{et al.}\xspace}

\newcommand{\ie}{\textit{i.e.,}\xspace}
\newcommand{\eg}{\textit{e.g.,}\xspace}

\mathchardef\mhyphen="2D

\renewcommand{\paragraph}[1]{\medskip \noindent \textbf{#1.\ }}

\graphicspath{ {figures/} }

\newcommand{\provider}{service provider\xspace}
\newcommand{\providers}{service providers\xspace}

\newcommand{\framework}{\textsl{\mbox{AuthStore}}\xspace}
\newcommand{\protocol}{\textsl{\mbox{CompactPAKE}}\xspace}

\newcommand{\cmark}{\ding{51}}
\newcommand{\xmark}{{\color{red} \ding{55}}}

\begin{document}

\title{AuthStore: Password-based Authentication and Encrypted Data Storage in Untrusted Environments}

\author{\IEEEauthorblockN{Clemens~Zeidler}
\IEEEauthorblockA{\textit{The University of Auckland} \\
Auckland 1142, New Zealand \\
clemens.zeidler@auckland.ac.nz}
\and
\IEEEauthorblockN{Muhammad~Rizwan~Asghar}
\IEEEauthorblockA{\textit{The University of Auckland} \\
Auckland 1142, New Zealand \\
r.asghar@auckland.ac.nz}
}

\maketitle

\begin{abstract}
Passwords are widely used for client to server authentication as well as for encrypting data stored in untrusted environments, such as cloud storage.
Both, authentication and encrypted cloud storage, are usually discussed in isolation.
In this work, we propose \framework, a flexible authentication framework that allows users to securely reuse passwords for authentication as well as for encrypted cloud storage at a single or multiple service providers.
Users can configure how secure passwords are protected using password stretching techniques.
We present a compact password-authenticated key exchange protocol (CompactPAKE) that integrates the retrieval of password stretching parameters. 
A {\em parameter attack} is described and we show how existing solutions suffer from this attack.
Furthermore, we introduce a password manager that supports CompactPAKE.

\end{abstract}


\section{Introduction}
\label{sec:Intro}

For authentication as well as for protecting digital assets, passwords are a simple, usable, and commonly used solution~\cite{Yee-2006-Passpet,VanAcker-2017-MeasuringLoginSecurity,Wright-2003-NCcryptfs,Kher-2005-SecuringDistributedStorage,Zeidler-2017-PortableCloud}.
The ubiquitously use of passwords encourages the reuse of passwords for multiple services~\cite{Florencio-2007-WebPasswordHabits,Das-2014-PasswordReuse,Ion-2015-ComparingSecurityPractices}.
However, this has important security and privacy implications since a leaked password gives an attacker access to multiple services, \eg to impersonate a user or to decrypt password-protected confidential user data.

Service providers often fail to store user passwords securely, \eg passwords are stored in plain text or weakly protected~\cite{Bauman-2015-PlainTextPasswordStorage}.
In general, it is unclear for users if and how service providers keep user passwords confidentially and, even worse, a service provider may itself be malicious and use a user password to impersonate a user at other service providers.
For instance, a web service provider can readout the user passwords straight from the login page by recording keystrokes~\cite{Ross-2005-PasswordAuthBrowserExtension} and so gain knowledge of all passwords entered by the user.

Passwords are not only used for authentication but also to encrypt data that can then be stored in potentially untrusted environments~\cite{Kher-2005-SecuringDistributedStorage,Zeidler-2017-PortableCloud}.
Here, it is important to remark that password-based authentication protocols should not leak the password to ensure that storage providers are unable to recover data that is encrypted with the same password~\cite{Boyen-2009-CredentialRetrievalReusablePassord,VanLaer-2016-HardenZeroKnowledge}.
Unfortunately, the discussion of a secure password-based authentication schemes is often neglected~\cite{Ferretti-2014-AccessToEncryptedCloud,Zarandioon-2012-K2C,Kher-2005-SecuringDistributedStorage,Zeidler-2017-PortableCloud,Zhao-2014-TowardCloudBasedPM,Jammalamadaka-2005-Pvault}, which can compromise secure data encryption.

For example, BoxCryptor\footnote{\url{www.boxcryptor.com}} supports a set of cloud storage providers as a storage back end.
If the same password is used for BoxCryptor and for the storage providers a malicious storage provider can decrypt the stored data.
Furthermore, encrypted storage providers, such as BoxCryptor or Mega\footnote{\url{mega.nz}}, let users authenticate on their web page with the encryption password.
This makes it possible for them to learn the plain password straight from their login page~\cite{Ross-2005-PasswordAuthBrowserExtension}.

A solution is that a user can choose two different passwords: one for authentication and one for data encryption~\cite{Leibenger-2013-StorageEfficientSVN}.
However, this solution suffers from usability concerns as it requires users to memorise two different passwords.
Identity and access delegation solutions, such as OpenID Connect\footnote{\url{openid.net/connect}} or OAuth\footnote{\url{oauth.net}}, avoid the use of passwords by using a trusted identity or authorisation server.
However, this approach only shifts the problem since an attacker who gains access to the authorisation server or the server itself may learn the user password.

In this paper, we present a novel framework called {\em \framework}.
\framework supports secure password-based authentication and ensures that the user password is not revealed to the service provider. 
The user password is secured from offline dictionary attacks~\cite{Narayanan-2005-FastDictionaryAttacks} by using key stretching techniques~\cite{Forler-2014-KDFOverview}.
By storing all key stretching parameters directly at the service provider our solution does not require any extra server and only username and password are required for a successful authentication.
The user is in full control of how secure the password is protected and stored at the service provider.
Furthermore, for service providers that offer data storage, \framework integrates secure password-based data encryption.
Our design makes it possible to securely reuse a single password for both authentication and privacy-preserving data storage at multiple service providers.

There are many password-based authentication protocols~\cite{Wu-1998-SecureRemotePasswordProtocol,Schnorr-1989-IdSmartCards,Gentry-2006-PAKE} but none of them assume the exchange of key stretching parameters.
We introduce a {\em parameter attack} that a service provider can perform during the user authentication with the goal to void the key stretching mechanism in order to make offline dictionary attacks easier.
We show how existing solutions, which assume strengthen passwords in the authentication protocol, suffer from parameter attacks.

We present the \protocol protocol that includes the exchange of key stretching parameters and is resistant against parameter attacks.
\protocol is based on a proven secure Password-Authenticated Key Exchange (PAKE) protocol but requires only four message passes between a user client and a \provider~\cite{Gentry-2006-PAKE,Bellare-2000-AuthenticatedKeyExchange}.
\protocol is an asymmetric authentication protocol, which means an attacker who compromises the service provider is not able to impersonate the user.

In a case study, we present a flexible \framework-based password manager that allows the secure storage of credentials in the cloud~\cite{Gasti-2012-SecurityPMDatabaseFormats,Li-2014-SecurityAnalysisWebBasedPM}.
This makes our solution not only useful for service providers that use \protocol but also for securely storing credentials for conventional password protocols.
We show how our password manager can speed up registration and authentication at service providers that support \protocol.
Furthermore, even without the password manager users are able to authenticate at \providers that uses \protocol.

Our contributions can be summarised as follows:
\begin{enumerate}[noitemsep]

	\item We propose \framework, a framework for secure single password authentication and password-protected, privacy-preserving cloud storage at a single or multiple service providers.


	\item We present \protocol, an asymmetric PAKE protocol with key stretching parameter retrieval that is secure against parameter attacks and only uses four message passes between a client and a service provider.

	\item A password manager that uses \framework to securely store arbitrary credentials in the cloud and that makes registration and authentication at \providers that support \protocol faster and easier.
	
\end{enumerate}

\section{System Requirements}
\label{sec:Requirements}
In the following, we define some core system requirements:

\paragraph{Simplicity}
The only information required by a user to authenticate and to register at a \provider is her username and password.
No other mechanism or service, such as a trusted key server, is required by the user.

\paragraph{Secure Password Storage}
The user must be able to configure how much an authentication password is strengthened before it is stored at a \provider, \ie the user is in control of how strong a password is secured from offline dictionary attacks.

\paragraph{Secure Authentication}
The used authentication protocol must ensure that no information about the password is leaked to any party, \ie dictionary attacks are impossible.
This includes that a \provider should not be able to learn any information about the entered password, \eg a mistakenly entered password used at a different service provider.

An attacker who compromises a service provider must not be able to impersonate the user using the stored authentication data.

\paragraph{Data Storage}
If a \provider supports data storage, a user must be able to securely reuse the authentication password to protect data without the risk that the \provider can access the data.

\section{System Model}
\label{sec:SystemModel}
In our system model, we have two main entities, \ie the {\em user} and the {\em service provider}.
A user owns one or more user accounts at one or more service providers.
The user communicates with a service provider through a trusted client software.
A user can authenticate at a service provider using her {\em username} and {\em user password}.
The service provider is responsible to manage a set of user accounts and stores all necessary parameters that are needed for the user authentication.
Optionally, the service provider can offer data storage to authenticated users.
The user is responsible to protect the data, \eg encrypt data, through the client.

\paragraph{Threat Model}
We assume that the service provider might be malicious.
The service provider may try to learn the user's clear text password in order to decrypt data that is encrypted with the same password or to impersonate the user at a different service provider.

In our approach, we use key stretching to protect the user password~\cite{Forler-2014-KDFOverview}.
We assume that the user chooses a user password with reasonable strength~\cite{Kelley-2012-PasswordStrength,VanAcker-2017-MeasuringLoginSecurity,Yampolskiy-2006-PasswordSelectionBehavior,Yan-2004-PasswordMemorability} and strong key stretching parameters so that it becomes practically infeasible for an attacker to recover the user password using an offline dictionary attack.

External attackers may learn or guess the username and try to impersonate the user.
The attacker may try to perform an online attack, guessing the user password.
The attacker may perform a Man-in-the-Middle (MitM) attack to monitor the authentication protocol.
Moreover, the attacker may gain access to the service provider in which case the attacker is treated the same as the \provider.

\paragraph{Our Approach}
In this work, we propose a framework for password-based authentication and data encryption where service providers are not necessarily trusted.
In order to strengthen the user password, we use a Key Derivation Function (KDF) to derive a strong {\em base key}.
In general, a KDF takes a set of parameters that determine how strong a derived key is protected from dictionary attacks.
The KDF parameters can be chosen by the user to control the trade-off between authentication time (KDF evaluation at the client) and password protection against possible attacks.

We use a simple but flexible method to derive an arbitrary number of {\em user keys} from a base key.
User keys can then be used for authentication, data encryption, or other purposes.
This also means that a base key can safely be reused for multiple purposes, even across multiple service providers.
As a result, the client can cache computationally expensive KDF evaluations and then quickly derive user keys from the cached values. 
All required parameters to derive a user key are stored at the service provider.

The choice of a suitable authentication protocol is crucial when using key stretching techniques.
More specifically, even if the password is strengthened, the key has to be treated as a weak key in the authentication protocol (see more discussion in Section~\ref{sec:SecurityAnalysis}).
In our approach, we use an asymmetric Password Authenticated Key Exchange (PAKE) protocol~\cite{Gentry-2006-PAKE} that has the required properties.
First, neither service providers nor external attackers can learn any information about the password used by the user.
The service provider only learns whether the user has provided the correct password or not.
This also means that the service provider cannot learn about erroneously entered passwords.
Second, a MitM attacker cannot use the intercepted information for an offline dictionary attack but is constrained to mount an online attack.
Third, if authentication data is stolen, it is insufficient to impersonate the user.

To support secure data storage we derive a user key and use this key to further protect arbitrary data encryption keys.
This provides the required bridge from secure authentication to secure cloud storage.
Since cloud storage can be highly domain specific, we refer to existing work for concrete solutions~\cite{Wright-2003-NCcryptfs,Kher-2005-SecuringDistributedStorage,Zeidler-2017-PortableCloud}.

\section{Solution Details}
\label{sec:Details}
In the following, we describe how a user can generate a strong base key using key stretching (Section~\ref{sec:KeyStretching}).
From this base key, a set of user keys can be generated (Section~\ref{sec:UserKeyGeneration}).
User keys are then used for authentication (Section~\ref{sec:RegAndAuth}) and for secure data storage (Section~\ref{sec:Storage}).
Moreover, we explain how authentication credentials can be updated (Section~\ref{sec:CredentialUpdate} and how an account is reset (Section~\ref{sec:AccountReset}).
We then analyse the security of our solution (Section~\ref{sec:SecurityAnalysis}).

\subsection{Password Strengthening}\label{sec:KeyStretching}
To strengthen the user password, a KDF is used~\cite{Kaliski-2000-PBKDF,Percival-2009-Scrypt,Biryukov-2016-Argon2,Forler-2014-KDFOverview}.
In general, a KDF takes the password $pw$, a salt value, and some cost parameters as input and returns a derive key $K$.
The cost parameters determine how expensive it is to evaluate a KDF.
For example, PDKDF2 uses an iteration count~\cite{Kaliski-2000-PBKDF} while more recent KDFs such as Argon2~\cite{Biryukov-2016-Argon2} allow users to specify memory requirements to hamper the use of fast GPUs and specialised hardware.
In the following, we refer to salt and cost parameters as $P_{K}$, which leads to the general definition of a KDF:
\[
K = KDF(P_{K},pw).
\]
where $K$ is the derived base key.




\begin{figure}[ht]
\centering
\includegraphics[width=\linewidth]{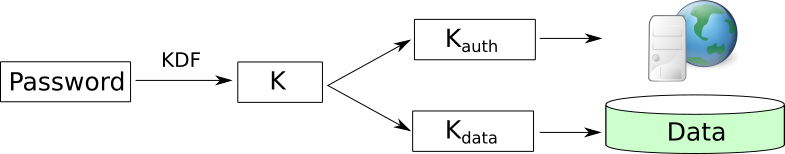}
\caption{Using a KDF, the user derives a base key $K$ from the user password.
From the base key, an arbitrary number of user keys can be derived.
These user keys can be used for authentication $K_{auth}$ and data encryption $K_{data}$.
}
\label{fig:KeyDerivationLoginKeyStore}
\end{figure}

\subsection{User Key Generation}
\label{sec:UserKeyGeneration}
To derive an arbitrary number of strong user keys from a single password using a single KDF evaluation, we use the following simple mechanism (see Figure~\ref{fig:KeyDerivationLoginKeyStore}).
As a basis for a user key $K_u$, a base key $K$ is required.
For each user key, a random salt value is generated and $K_{u}$ is derived as follows:
\[
K_{u} = H(K, salt) = H(KDF(P_K,pw), salt) = U(P_u,pw)
\]
with a cryptographic one-way hash function $H$ and the {\em user key parameters} $P_{u} = [P_{K}, salt]$.

\subsection{Registration and Authentication}
\label{sec:RegAndAuth}
For the purpose of authentication at a service provider, the user derives a user key $K_{auth}$ from the user password.
For brevity reason, it is referred to as: $\pi = K_{auth}$.
We use an asymmetric PAKE protocol for authentication~\cite{Gentry-2006-PAKE}.
For this protocol, the user requires the authentication key $\pi$ and the service provider requires some verification value $V_{\pi}$ to verify the users' knowledge of $\pi$.
In general, $V_{\pi}$ can be a set of values.
In Section~\ref{sec:CompactPAKE}, we describe a compact variant of an asymmetric PAKE protocol and describe in detail how the verification value $V_{\pi}$ is derived from $\pi$.

For the registration of a user account, the user chooses the user key parameters $P_{\pi}$, derives $V_{\pi}$ (by first deriving $\pi$), and uploads $P_{\pi}$ and $V_{\pi}$ to the \provider.
To authenticate, the user requests $P_{\pi}$ from the service provider and derives the login key $\pi$, which is then used to authenticate using the PAKE protocol.
At a successful authentication, the PAKE protocol produces a mutual secure session key $sk$ which can be used to protect further communication between client and \provider~\cite{Bellare-2000-AuthenticatedKeyExchange}.
In the following, we assume a secure connection after authentication.

\subsection{Secure Data Storage}
\label{sec:Storage}
If the \framework \provider offers data storage, the user may ensure data confidentiality by employing encryption techniques~\cite{Wright-2003-NCcryptfs,Kher-2005-SecuringDistributedStorage,Zeidler-2017-PortableCloud}.
We show a simple but flexible approach of how data can be protected using the same password as used for authentication.

For the encryption, the user generates some user key parameters $P_{u,data}$ and derives a new user key $K_{data}$.
To minimise the computational cost, the base key parameters $P_{K}$ that are used to derive the authentication key $\pi$ can be reused, \ie the base key has already been derived for the authentication.

Instead of encrypting data directly with $K_{data}$, we propose a more flexible key chain approach\footnote{\url{gitlab.com/groups/cryptsetup}}.
This approach has the advantage that multiple passwords can be supported and passwords can be changed without the need to re-encrypt the data~\cite{Zeidler-2017-PortableCloud}.
Here, a random symmetric encryption key $k_{sym}$ is generated and used for data encryption.
$k_{sym}$ is then encrypted with $K_{data}$.
We call the set of the encrypted key $k_{sym}$ and $P_{u,data}$ the data parameters $P_{data} = [enc(k_{sym}), P_{u,data}]$.
The encrypted data and $P_{data}$ can securely be stored at a \provider.

\subsection{Password Change and Key Update}\label{sec:CredentialUpdate}
To change the user password, a user chooses new user key parameters $P'_{\pi}$ and derives $V'_{\pi}$ from the new password.
At a successful authentication with the old password, $P'_{\pi}$ and $V'_{\pi}$ are updated at the \provider.
The same approach can be used to just update the user key parameters.
This might be desirable if more secure KDF parameters should be used while reusing the existing password.

If encrypted data is stored at the \provider, a new encryption key $K_{data}$ is generated, $k_{sym}$ is re-encrypted with $K_{data}$, and $P_{data}$ is updated at the \provider.

\subsection{Account Reset}
\label{sec:AccountReset}
In case the user forgets her username or password used at a certain \provider, traditional recovery methods can be applied.
For example, the user can request an account reset through a trusted third party channel such as an ordinary email account.
In response, the service provider generates a temporary random authentication key $\pi'$ and the matching verification value $V_{\pi'}$ and sends the key $\pi'$ to the user.
The user uses this authentication key to authenticate using the normal protocol.
Here the authentication key derivation step is ignored since the authentication key $\pi'$ is already known to the user.
After a successful authentication, the user chooses a new password, as described above.

Recovering data that is protected with the lost user password would require techniques such as secret sharing~\cite{Camenisch-2014-MementoSinglePassword} or a local key backup. However, this is incompatible with our simplicity requirement.


\subsection{Security Analysis} 
\label{sec:SecurityAnalysis}
In this section, we analyse the security of \framework.
Moreover, we describe a novel parameter attack that targets authentication schemes that include password stretching.

\paragraph{Key Derivation}
In our approach, we use a cryptographic one-way function to derive multiple user keys from a single base key.
This means just a leaked user key $K'_u$ cannot be used to learn information about the base key or other user keys.
However, if the matching $P'_u$ is as well known, an attacker can perform an offline dictionary attack.

\paragraph{Offline Dictionary Attack}
Offline dictionary attacks can always be performed by the \provider, \ie the \provider can fully simulate the authentication protocol.
Our approach fully relies on the assumption that the user password is reasonably strong~\cite{Kelley-2012-PasswordStrength,VanAcker-2017-MeasuringLoginSecurity,Yampolskiy-2006-PasswordSelectionBehavior,Yan-2004-PasswordMemorability,Ur-2017-DesignPasswordMeter} so that it can sufficiently be strengthened using a KDF~\cite{Forler-2014-KDFOverview}.
%
Because of the used PAKE protocol, an external attacker can only mount an offline dictionary attack if either the service provider actively leaks $P_{\pi}$ and $V_{\pi}$ or the attacker is able to compromise the service provider.

\paragraph{Password Update and Parameter Ageing}
At any point, the user may want to change to a more secure password or use more secure base key parameters.
The user has to be aware that there is no way to force the \provider to discard old $P_{\pi}$ and $V_{\pi}$ values.
Thus, a malicious \provider may choose the weakest available user key parameters for a dictionary attack.

\begin{figure*}[t]
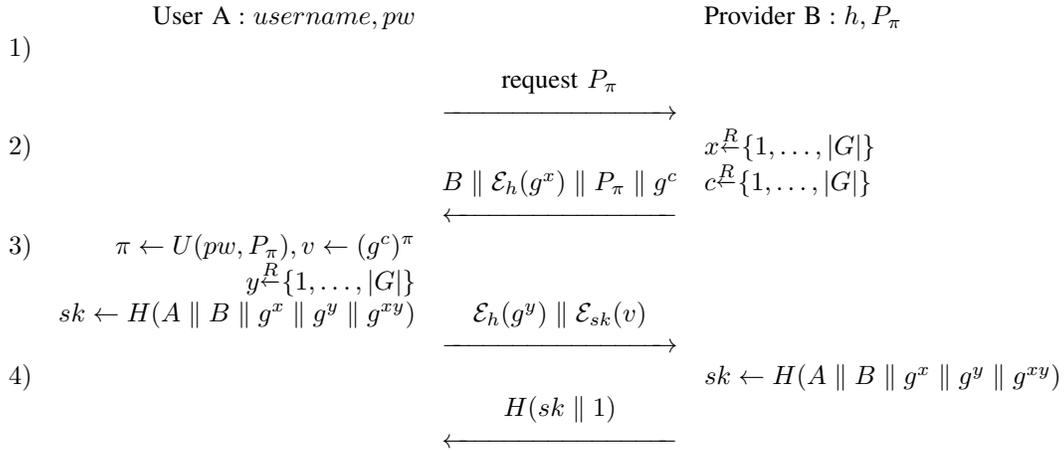

\[
\begin{array}{r r c l}
&\mbox{User A}: username, pw 		&								&\mbox{Provider B}: h, P_{\pi}	\\
1)&			&													&				\\
&			&\mbox{request $P_{\pi}$}								&				\\
&			&$\rightarrowfill$									&				\\
2)&	 		&													&x^{\underleftarrow{R}}\{1, \ldots, |G|\}\\
&	 		&B\parallel\mathcal{E}_{h}(g^x)\parallel P_{\pi}\parallel	g^c	&c^{\underleftarrow{R}}\{1, \ldots, |G|\}\\
&			&$\leftarrowfill$									&				\\
3)&\pi \leftarrow U(pw, P_{\pi}), v \leftarrow (g^c)^{\pi}&									&				\\				
&y ^{\underleftarrow{R}} \{1, \ldots, |G|\}&							&				\\
&sk \leftarrow H(A\parallel B\parallel g^x\parallel g^y\parallel g^{xy})&\mathcal{E}_{h}(g^y)\parallel \mathcal{E}_{sk}(v)&\\
&			&$\rightarrowfill$									&				\\	
4)&			&													& sk \leftarrow H(A\parallel B\parallel g^x\parallel g^y\parallel g^{xy})\\
&			&H(sk\parallel 1)									&\\
&			&$\leftarrowfill$									&\\
\end{array}
\]
\caption{The CompactPAKE protocol.}
\label{fig:CompactPAKE}
\end{figure*}

User key parameters, which deemed to be secure at the time of account creation may ``age'' over time and become insecure with the emergence of more powerful hardware.
This means accounts can become vulnerable over time.
Data encrypted with an encryption key that is based on an outdated base key should be re-encrypted.
However, as before, a malicious service provider can still mount an attack on an older version of the encrypted data that is encrypted using weaker user key parameters.
For this reason, data that should be stored long-term should be encrypted using a high-entropy password~\cite{Kelley-2012-PasswordStrength,VanAcker-2017-MeasuringLoginSecurity,Yampolskiy-2006-PasswordSelectionBehavior,Yan-2004-PasswordMemorability,Ur-2017-DesignPasswordMeter}.

\paragraph{Parameter Attack}
In a parameter attack, the service provider reports tempered, weak, authentication parameters $P'_K$ when requested by the user.
In this way, the user is tricked to authenticate with a weak authentication key $\pi'$.
The attack is applicable if the service provider is able to learn information about $\pi'$ during the authentication protocol.
In this case, the service provider may perform a dictionary attack on $\pi'$. 

For example, the Secure Remote Password (SRP) protocol~\cite{Wu-1998-SecureRemotePasswordProtocol} is vulnerable to parameter attack when used in our scenario.
In this protocol, the client sends a value $M_1=H(A (g^{\pi'})^u)^b$ to the \provider. 
Here, $H$ is a one-way hash function and $A, u, b$ parameters known to the \provider.
When, due to the parameter attack, $\pi'$ is a low-entropy key, this equation can be solved by the \provider using a dictionary attack.
Similarly, the Schnorr protocol~\cite{Schnorr-1989-IdSmartCards} requires a high entropy secret.
A parameter attack can be avoided by protocols designed for low entropy authentication keys~\cite{Bellare-2000-AuthenticatedKeyExchange,Gentry-2006-PAKE}.

This makes approaches that use key stretching to derive a strong authentication key while not requiring an authentication protocol that assumes a weak authentication key vulnerable.
For example, the password manager Passpet uses SRP~\cite{Yee-2006-Passpet} while Van Laer at al.~\cite{VanLaer-2016-HardenZeroKnowledge} are using the Schnorr protocol.

\section{Compact PAKE Protocol}\label{sec:CompactPAKE}

An asymmetric PAKE protocol usually requires the exchange of six messages between the client and the \provider~\cite{Gentry-2006-PAKE,Jablon-1997-ExtendedPasswordKeyExchange}.
To derive the login key $\pi$, the user first has to request the authentication parameters $P_{\pi}$ from the service provider which adds two additional messages to the PAKE protocol.
In the following, we propose \protocol, a PAKE protocol that only requires four messages including the retrieval of $P_{\pi}$.
Our approach is based on existing proven secure building blocks, \ie the symmetric PAKE protocol EKE2~\cite{Bellare-2000-AuthenticatedKeyExchange} and a simple authentication scheme used for the B-Speke protocol~\cite{Jablon-1997-ExtendedPasswordKeyExchange}.

\paragraph{Registration}
For the registration, the user generates an authentication key $\pi$ from some chosen authentication parameters $P_{\pi}$.
The user chooses a cyclic group generator $g$ and calculates $h = g^{\pi}$, which is used as the verification value $V_{\pi}$.
The user deploys $h$ and the authentication parameters $P_{\pi}$ at the service provider.

\paragraph{Authentication}
\protocol closely resembles the symmetric EKE2 protocol~\cite{Bellare-2000-AuthenticatedKeyExchange}, where the user $A$ is the verifier and the \provider $B$ is the prover.
The protocol (Figure~\ref{fig:CompactPAKE}) works as follows.

\begin{enumerate}[noitemsep]
	\item The user requests the authentication parameters $P_{\pi}$.
	
	\item The \provider generates the random values $x$ and $c$ and sends back the \provider id $B$, $\mathcal{E}_{h}(g^x)$, $P_{\pi}$ and $g^c$ (with $\mathcal{E}_{h}$ an encryption function).
	
	\item The user derives $\pi$, calculates $v = (g^c)^{\pi}$, generates a random value $y$ and the session key
$sk=H(A\parallel B\parallel g^x\parallel g^y\parallel g^{xy})$ is calculated. 
	The user responses to the service provider with $\mathcal{E}_{h}(g^y)$ and $\mathcal{E}_{sk}(v)$. 
	
	\item The service provider derives the session key $sk=H(A\parallel B\parallel g^x\parallel g^y\parallel g^{xy})$ and uses $sk$ to decrypt $v$. 
	If $v$ is not equal $h^c$ the protocol is terminated. 
	Otherwise, the user is authenticated and the service provider responds with $H(sk\parallel 1)$.
	
	\item The user verifies that the user's version of $H(sk\parallel 1)$ matches the received value from the \provider. 
	If the values do not match, the protocol is terminated. 
	
\end{enumerate}

\paragraph{Correctness}
Concerning user authentication, only Step 4 differs from EKE2~\cite{Bellare-2000-AuthenticatedKeyExchange}.
This step is correct since $v = (g^c)^{\pi} = h^{c}$.

\paragraph{Security Analysis}
There are two main differences to the EKE2 protocol~\cite{Bellare-2000-AuthenticatedKeyExchange}.
First, because the user starts with the parameter request, the EKE2 part is started by the service provider rather by the user.
Second, the user is authenticated with an asymmetric approach.
In the following, we analyse the differences in detail.

In Step 1, the user solely requests the login parameters $P_{\pi}$.
In Step 2 the service provider sends $B\parallel\mathcal{E}_{h}(g^x)$, which is identical to the first step of EKE2.
Furthermore, the service provider sends $P_{\pi}$ and the random challenge $g^c$, which both do not reveal any sensitive information to an attacker.

In Step 3, the user sends $\mathcal{E}_{h}(g^y)$, which is again the same as in EKE2.
However, differently to EKE2, the user sends $\mathcal{E}_{sk}(v)$.
From EKE2, it is known that the session key $sk$ can only be calculated correctly by both parties if both parties know $h$.
This means the service provider learns the correct $v$ only if $h$ is known.
Since $v = (g^c)^{\pi} = (g^{\pi})^c = h^c$, the service provider does not learn any new information other than the fact that the user not only knows $h$ but also $\pi$.
Step 4 is again identical to EKE2.

\section{Password Manager}
\label{sec:PasswordManager}

To provide a complete authentication solution and to demonstrate the flexibility of \framework, we present a password manager that can securely store arbitrary credentials (\eg web login passwords) at any \framework provider.

Password managers have been widely discussed~\cite{Jammalamadaka-2005-Pvault,Camenisch-2014-MementoSinglePassword,Yee-2006-Passpet,Halderman-2005-SecurelyManagingPasswords,Fatma-2017-AutoPass} for which reason we focus on the management of \protocol user keys and how this can speed up the registration and authentication at \framework \providers.

\subsection{Design}
The design of the proposed password manager is as follows.
To securely store key material the password manager uses password protected encrypted storage as described in Section~\ref{sec:Storage}.

The encrypted password manager as well as encryption parameters $P_{data}$ (Section~\ref{sec:Storage}) are kept and managed locally but can be synchronised to an \framework \provider, \ie the encrypted password manager and $P_{data}$ are stored at the \framework \provider.

\begin{figure}[ht]
\centering
\includegraphics[width=0.7\linewidth]{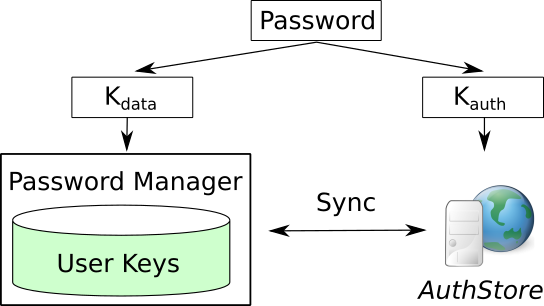}
\caption{A single password is used to access to the encrypted password manager and for authentication with an \framework \provider.
The \provider can not leverage the authentication process to gain access to the stored data.
}
\label{fig:PasswordManager}
\end{figure}

Using this approach users can login to the \framework \provider and access the password manager with the same password (see Figure~\ref{fig:PasswordManager}).

\subsection{Implementation}
We implemented our prototype as a browser extension to allow secure web authentication.
Note that an implementation that is loaded with a web page cannot be trusted since the service provider can manipulate the provided code~\cite{Ross-2005-PasswordAuthBrowserExtension}.
Thus, a trusted client-side implementation is necessary.

Our implementation can be used to store general web login passwords as well as \provider user keys.
Furthermore, the browser extension offers support to authenticate at arbitrary \framework \providers.
This can be used to remove the hard task from web developer to implement web authentication pages securely~\cite{VanAcker-2017-MeasuringLoginSecurity}, \ie our browser extension can be leveraged for user authentication.
The source code of our implementation can be accessed on\footnote{\url{gitlab.com/czeidler/authstore}}.

\subsection{Discussion}
In the following we discuss the advantages of the proposed password manager.

\paragraph{Avoiding expensive key derivation}
The derivation of a user key from a password can, depending on the chosen base key parameters, take a considerable time.
One way to mitigate this problem is to reuse base keys for multiple \providers.
However, a more general solution is to store user keys in the proposed password manager.
Thus, user keys derived from different base keys can be cached in the password manager.

Note, that no base keys should be stored in the password manager.
This prevents that even when the content of the password manager is inadvertently exposed to an attacker, other user keys which are not stored in the password manager stay secure.
Furthermore, the base key can continued to be used for future user keys.

\paragraph{Password-less registration}
The password manager can simplify the user registration at new \framework \providers by removing the need for letting users choose a new authentication password but only requiring users to enter a new username.
For this to work, the password manager uses existing base key parameters and automatically derive a new authentication key.
Reusing previously derived base keys can avoid the expensive KDF evaluation, \eg by using the base key parameters used to unlock the password manager.
By proposing a previously used username to the user, the registration can be further simplified and becomes as easy as clicking a ``Register'' button.

\paragraph{Resilient against lose of the password manager}
Another advantage of our approach is that reliable storage of the password manager is not a hard requirement, \ie the user remains able to authenticate at \framework \providers even without the password manager.
This is especially useful if the user does not own a cloud storage account at a \provider and only stores the password manager locally.
While the password manager makes authentication more efficient, the user only needs to remember her username and password to authenticate at a \provider using \protocol, \ie no other information from the password manager is needed.
This obviously fails when the set of passwords becomes too large to remember.
However, \protocol makes single password usage secure and the ability to easily change a password (Section~\ref{sec:CredentialUpdate}) makes a single password scenario possible. 

\section{Related Work}
\label{sec:RelatedWork}

Another group of authentication protocols leverage the use of external devices, such as mobile phones~\cite{Sun-2012-oPassAuthProtocolPWStealing,Jarecki-2016-DeviceEnhancedPWProtocol}, or uses a credential server~\cite{Boyen-2009-CredentialRetrievalReusablePassord}.

\begin{table*}[t]
\centering
\caption{Comparative analysis of different shcemes. \label{tab:Comparison}}
\begin{tabular}{ |m{15em}|c|c|c|c| }
 \hline
\textbf{Scheme} & \textbf{Simplicity} & \textbf{Secure Password Storage} & \textbf{Secure Authentication} & \textbf{Data Storage} \\
 \hline
Password~hashing~\cite{Ross-2005-PasswordAuthBrowserExtension} &  \cmark & \xmark (dictionary attacks) & \xmark & - \\ \hline

Strengthen password hashing~\cite{Halderman-2005-SecurelyManagingPasswords} &  \cmark & \xmark(fixed parameters) & \xmark & - \\ \hline

HPAKE~\cite{Boyen-2007-HaltingPasswordPuzzles,Blocki-2016-ClientCashProtectingPasswords} & \cmark & \cmark & \cmark & - \\ \hline

Secret sharing~\cite{Camenisch-2014-MementoSinglePassword} & \xmark (setup requires n servers) & \cmark & \cmark & - \\ \hline

Decoy passwords~\cite{Bojinov-2010-KamouflageLossResistantPM,Chatterjee-2015-PWVaultsNaturalLanguage}& \cmark & \xmark (see~\cite{Golla-2016-SecurityOfResitantPWVaults}) & - (not discussed) & -\\ \hline

Physical devices~\cite{McCarney-2012-TapasPW,Horsch-2015-Palpas,Shirvanian-2017-SphinxPWStore, Horsch-2017-UpdateTolerantPWBackup} & \xmark (requires physical device) & \cmark & \cmark (relies on physical device) & \cmark \\ \hline


Pvault~\cite{Jammalamadaka-2005-Pvault} & \cmark & \cmark & \xmark & \xmark\\ \hline
Cloud-based Password Manager~\cite{Zhao-2014-TowardCloudBasedPM} & \cmark & \cmark & \xmark & \xmark\\ \hline

Passpet~\cite{Yee-2006-Passpet} & \cmark & \cmark & \xmark (parameter attack) & \xmark\\ \hline
{\bf \framework} & \cmark & \cmark & \cmark & \cmark \\ \hline
\end{tabular}
\vspace{0.1cm}
\end{table*}

  




There is a lot of work on password-based authentication protocols.
While Wu \etal assume a strong password~\cite{Wu-1998-SecureRemotePasswordProtocol}, more recent protocols such as PAKE protocols can work with weak passwords in order to prevent attackers to perform dictionary attacks on the exchanged information~\cite{Bellovin-1992-EncryptedKeyExchange,Bellare-2000-AuthenticatedKeyExchange,Gentry-2006-PAKE}.
Asymmetric versions of PAKE protocols prevent an attacker to impersonate a user when compromising a service provider~\cite{Bellovin-1993-AugmentedEKE,Jablon-1997-ExtendedPasswordKeyExchange,Gentry-2006-PAKE}.
The proposed CompactPAKE protocol is closely related to EKE2~\cite{Bellare-2000-AuthenticatedKeyExchange} but is more compact, includes user key parameter retrieval, and asymmetric authentication~\cite{Gentry-2006-PAKE}.











BetterAuth uses a PAKE scheme for authentication but it does not cover password stretching~\cite{Johns-2012-BetterAuthWebAuthRevisited}.
Similar to our work, Van Laer \etal propose to use a KDF for password stretching and to store the required salt parameter at the service provider~\cite{VanLaer-2016-HardenZeroKnowledge}.
Compared to our work, they only use pre-defined KDF parameters, such as CPU cost, and only have configurable salt, which makes their approach inflexible regarding account ageing (see Section~\ref{sec:Details}).
Furthermore, they use the Schnorr protocol~\cite{Schnorr-1989-IdSmartCards}, which is vulnerable to parameter attacks, \eg when the provider returns a manipulated salt value.

An interesting KDF approach that is not vulnerable against parameter attacks is a Halting KDF (HKDF)~\cite{Boyen-2007-HaltingPasswordPuzzles,Blocki-2016-ClientCashProtectingPasswords}.
When evaluating an HKDF, the user executes a KDF algorithm till a halting parameter is encountered.
An invalid password can not be discarded with absolute certainty.
This results in more than 3 times more work for an attacker when performing a dictionary attack~\cite{Boyen-2007-HaltingPasswordPuzzles}.
The HPAKE authentication protocol uses an HKDF and stores the required halting parameter at the predefine provider~\cite{Boyen-2009-Hpake}.
The halting parameters are concealed from external attackers using a hidden credential retrieval scheme~\cite{Boyen-2009-CredentialRetrievalReusablePassord} to make offline dictionary attacks for external attackers impossible.
Using HPAKE, a parameter attack would not be feasible since the user's HKDF calculation would not terminate.
A disadvantage of HKDFs is the usability issue that a user, who accidentally entered a wrong password, does not get a timely feedback about the mistake.
Our approach works with any conventional state of the art KDFs.



Another group of authentication protocols leverage the use of external devices, such as mobile phones~\cite{Sun-2012-oPassAuthProtocolPWStealing,Jarecki-2016-DeviceEnhancedPWProtocol}, or uses a credential server~\cite{Boyen-2009-CredentialRetrievalReusablePassord}.
In our work, no extra device or server is needed for the authentication with the \provider.





\paragraph{Password Management}
A server-less approach that allows a user to reuse a password on multiple web sites is to derive an authentication key using a hash function that takes the domain name and the password as arguments~\cite{Ross-2005-PasswordAuthBrowserExtension}.
Since this approach is vulnerable to dictionary attacks, it has been proposed to apply the hash function $n$ times to strengthen the password~\cite{Halderman-2005-SecurelyManagingPasswords}.
However, $n$ is a hard-coded value, which means the approach cannot be adapted to future hardware.
Moreover, these approaches suffer from the problem that the site password can not be changed without changing the master password or to remember additional state information.

One way to store the password manager's master key (but not the password manager itself) is to leverage secret sharing~\cite{Camenisch-2014-MementoSinglePassword}.
This approach requires a set of storage servers from which at least a certain subset is not compromised by an attacker.

Decoy passwords (honeywords) or honey encryption can be used to protect passwords to make online attacks difficult~\cite{Bojinov-2010-KamouflageLossResistantPM,Chatterjee-2015-PWVaultsNaturalLanguage}.
However, recent work showed that real passwords can be distinguished from decoy passwords with high accuracy~\cite{Golla-2016-SecurityOfResitantPWVaults}

Some other approaches leverages user devices, such as mobile devices~\cite{McCarney-2012-TapasPW,Horsch-2015-Palpas,Shirvanian-2017-SphinxPWStore} or a smart card~\cite{Horsch-2017-UpdateTolerantPWBackup} to protect the user password.
A common issue is that losing the user devices also means to lose access to the password store and a local backup of the device needs to be maintained manually.

Pvault offers encrypted cloud based data storage and password management but uses the same password for data encryption and for authenticate at the storage server~\cite{Jammalamadaka-2005-Pvault}.
Zhao \etal use a KDF to protect a password store on secure reliable cloud storage but only a simple authentication method is assumed~\cite{Zhao-2014-TowardCloudBasedPM}, \ie the \provider must be trusted.
Passpet uses the same simple key stretching technique as used by Halderman \etal~\cite{Halderman-2005-SecurelyManagingPasswords} but stores KDF parameters on a Passpet server~\cite{Yee-2006-Passpet}.
However, for authentication, Passpet uses the Secure Remote Password protocol~\cite{Wu-1998-SecureRemotePasswordProtocol}, which is vulnerable to parameter attacks.

Table~\ref{tab:Comparison} shows which related work fulfils our requirements from Section~\ref{sec:Requirements}.
The data storage requirement is only evaluated for password managers that store passwords at a server.
Only \framework fulfils all requirements, \ie it provides simple password based login, passwords can arbitrarily be strengthen, the authentication protocol doesn't reveal any information about the used password, and the authentication password can securely be used for data storage.

\section{Conclusion}
\label{sec:Conclusion}

In this paper, we presented \framework, a secure password-based authentication method.
We showed that a strong authentication method that keeps the user password confidentially is a requirement for secure password-based encrypted cloud storage.
\framework allows users to securely reuse passwords for authentication at multiple service providers as well as for secure data encryption.
\framework only requires a single service, \ie a service provider, to operate.
Users only need to remember username and password to authenticate and access their encrypted data.
Using \framework, users are in control of how secure passwords are protected using key stretching.
We discussed a parameter attack and showed how other solutions are vulnerable to this attack.
We presented \protocol, a compact asymmetric PAKE protocol that includes the retrieval of key stretching parameters and requires fewer communication messages than other PAKE protocols.

In a case study, we presented a \framework-based password manager that allows users to securely store arbitrary credential such as web login passwords in the cloud.
We showed how the password manager helps to make user registration and authentication at service providers that support \protocol faster and more convenient.

\bibliographystyle{IEEEtran}
\bibliography{references}

\end{document}